\documentclass[twocolumn,aps,pra,preprintnumbers,showpacs,bibnotes,amsmath,amssymb]{revtex4}
\usepackage{graphicx}
\usepackage{multirow}
\usepackage{color}
\usepackage{units}
\usepackage{mathrsfs}
\usepackage[colorlinks,urlcolor=blue,citecolor=blue,linkcolor=blue]{hyperref}

\begin{document}

\title{Hyperfine structure of laser-cooling transitions in fermionic erbium-167}

\author{Albert Frisch}
\author{Kiyotaka Aikawa}
\author{Michael Mark}
\author{Francesca Ferlaino}
\affiliation{Institut f\"ur Experimentalphysik and Zentrum f\"ur Quantenphysik, Universit\"at Innsbruck, Technikerstra{\ss}e 25, 6020 Innsbruck, Austria}

\author{Ekaterina Berseneva}
\altaffiliation{Alternate address:  Division of Quantum Mechanics, St.Petersburg State University, 198904, Russia}
\affiliation{Department of Physics, Temple University, Philadelphia, Pennsylvania 19122, USA}
\author{Svetlana Kotochigova}
\affiliation{Department of Physics, Temple University, Philadelphia, Pennsylvania 19122, USA}

\date{\today}


\begin{abstract}
We have measured and analyzed the hyperfine structure of two lines, one at 583 nm and one at 401 nm, of the only stable fermionic isotope of atomic erbium as well as determined its
isotope shift relative to the four most-abundant bosonic isotopes.  Our work  focuses on the $J \rightarrow J +1$ laser cooling transitions from the  $[\rm{Xe}]4f^{12}6s^{2}
(^3H_6)$ ground state to two levels of the excited $[\rm{Xe}]4f^{12}6s6p$ configuration, which are of major interest for experiments on quantum degenerate dipolar Fermi gases.
From a fit to the observed spectra of the strong optical transition at $401$\,nm we find that the magnetic dipole and electric quadrupole hyperfine constants for the excited state
are $A_{e}/h=-100.1(3)\,\mathrm{MHz}$ and $B_{e}/h=-3079(30)\,\mathrm{MHz}$, respectively. The hyperfine spectrum of the narrow transition at $583$\,nm, was previously observed and
accurate $A_{e}$ and $B_{e}$ coefficients are available. A simulated spectrum based on these coefficients agrees well with our measurements.  We have also determined the hyperfine
constants using relativistic configuration-interaction {\it ab-initio} calculations. The agreement between the {\it ab initio} and fitted data for the ground state is better than
0.1\% , while for the two excited states the agreement is  1\% and 11\%  for the $A_e$ and $B_e$ constants, respectively.
\end{abstract}

\maketitle

\section{Introduction}
\label{intro}

The field of ultracold quantum gases has historically heavily relied on alkali-metal atoms. Only recently, the use of non-alkali-metal atoms has gained attention as a means to explore fascinating quantum phases of matter that are not accessible with alkali-metal species. Species with multiple unpaired valence electrons have rich atomic energy spectra and exhibit various types of coupling between the electronic angular momentum $\vec J$ and the nuclear spin $\vec I$ of the atom. For instance, fermionic alkaline-earth-metal atoms have $J\!=\!0$ and $I\!\ne\!0$ and the electronic and nuclear angular momenta decouple. This decoupling is at the center of proposals for efficient quantum simulation \cite{Daley2008qcw,Stock2008eog,Gorshkov2009aem} and quantum magnetism \cite{Hermele2009mio,Gorshkov2010tos,FossFeig2010hfi,Hung2011qmo}. Recently, degenerate Bose and Fermi gases of Ca \cite{Kraft2009bec}, Sr \cite{Stellmer2009bec,MartinezdeEscobar2009bec,Stellmer2011dam}, and the alkaline-earth-metal-like Yb atoms \cite{Takasu2003ssb, Fukuhara2007dfg} have been realized.  

Lanthanides with submerged 4f-shell electrons are a novel class of atoms that attract attention in the field of ultracold quantum physics. Lanthanide atoms can have an exceptionally large electronic angular momentum $\vec J$ resulting from the alignment of the angular momenta of the submerged electrons.  Consequently, these species can have strong magnetic moments $\mu$ as large as $10\,\mu_B$, where $\mu_B$ is the Bohr magneton. The mutual interaction is dominated by long-range magnetic dipole-dipole forces. Their dipolar character can be one hundred times larger than that for alkali-metal atoms. This key property makes lanthanides prime candidates for the study of atomic dipolar physics \cite{Baranov2008tpi,Lahaye2009tpo, Baranov2012cmt}. Dy \cite{Lu2011sdb,Lu2012qdd} and Er \cite{Aikawa2012bec}, with $\mu=10\,\mathrm{\mu_{B}}$ and $7\,\mathrm{\mu_{B}}$, respectively, have been recently brought to quantum degeneracy while others are under investigation \cite{Saffman2008stn, Sukachev2010mto,Sukachev2011lco}.  

The success of quantum-degenerate-gas experiments relies on a precise understanding of the atomic properties, such as energy levels, hyperfine structures, and atomic polarizabilities. However, for unconventional atomic species, such as lanthanides, the available knowledge is in many  instances insufficient for laser cooling and trapping purposes. Therefore, dedicated experiments need to be conducted en route
to quantum degeneracy \cite{Mcclelland2006lcw,Sukachev2010mto,Youn2010dmo, Dzuba2011dpa}.

In this paper, we present a combined experimental and theoretical investigation of the hyperfine structure of the only stable fermionic erbium isotope, $^{167}$Er. In particular, we obtain the magnetic dipole, $A$, and electric quadrupole, $B$, hyperfine structure constants  for the ground and two electronically excited states of $^{167}$Er, which are relevant for laser cooling experiments  \cite{Mcclelland2006lcw, Frisch2012nlm}. The two electronic excited states investigated are the one at a wavelength of $582.67\,\mathrm{nm}$ (corresponding to photon energy $E/(hc)=17157.307\,\mathrm{cm}^{-1}$) and one at $400.796\,\mathrm{nm}$ ($E/(hc)=24943.272\,\mathrm{cm}^{-1}$) from the ground state \cite{Erbium}. Here $h$ is Planck's constant and $c$ is the speed of light. In addition to the study of the hyperfine constants, we also obtained the isotope shift of $^{167}$Er  relative to the most-abundant bosonic isotopes. Our work provides important information for future experiments on quantum-degenerate Fermi gases of strongly dipolar Er atoms. 

In a previous work, we used the optical transitions at about $401\,\mathrm{nm}$ and $583\,\mathrm{nm}$ for Zeeman slowing (ZS) and magneto-optical trapping (MOT) applications \cite{Frisch2012nlm}. We demonstrated  efficient laser cooling for five Er isotopes, including the fermionic one. However, the realization of a MOT of fermionic Er isotope was challenging since only the hyperfine structure of the  ground and the $583\,\mathrm{nm}$-excited state were known \cite{Childs1983,Jin1990hsa}, while the one of the state at $401\,\mathrm{nm}$ was unknown prior to this work. To operate the Zeeman slower and the transversal cooling stage we had in fact to proceed empirically  and try different locking points for the light at $401\,\mathrm{nm}$ before being able to produce a MOT of Fermions.

\begin{figure}[b]
\includegraphics[width=0.9\columnwidth] {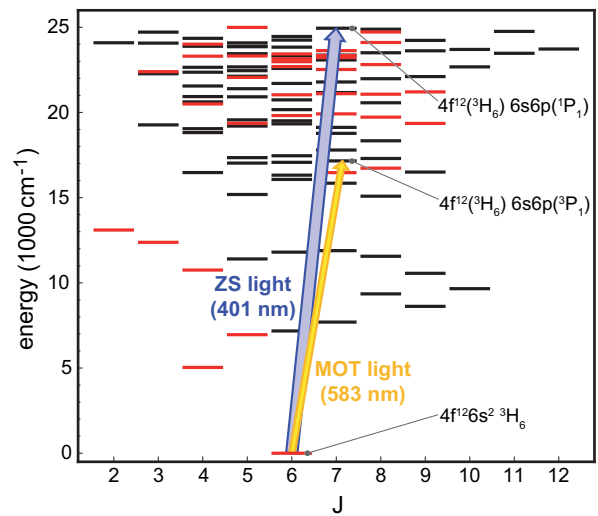}
\caption{(Color online) Energy levels of atomic Er up to  $E/(hc)=\unit[25 000]{cm^{-1}}$ for different electronic angular momentum quantum numbers $J$ \cite{Erbium,Ban2005lct}. States with odd (even) parity are indicated by black (red) horizontal lines. The two relevant laser-cooling transitions at $401\,\mathrm{nm}$ and $583\,\mathrm{nm}$ are indicated by arrows.}
\label{fig:levelscheme}
\end{figure}

Figure \ref{fig:levelscheme} shows the atomic level scheme of Er. The electronic ground state belongs to the $\rm [Xe]4f^{12}6s^{2}$ configuration and has a large orbital angular momentum quantum number $L=5$ (H state) and a total electronic angular momentum quantum number $J=6$. The excited states at $401\,\mathrm{nm}$ and $583\,\mathrm{nm}$ belong to the $\rm [Xe]4f^{12}6s6p$ configuration and have singlet $^1$P$_1$ and triplet $^3$P$_1$ character for the outer two valence electrons, respectively.  Both excited states have a total electron angular momentum $J=7$. 
Erbium has six stable isotopes with natural abundance being 33.6\,\% for $^{166}$Er, 26.8\,\% for $^{168}$Er, 23.0\,\% for $^{167}$Er , 14.9\,\% for $^{170}$Er, 1.61\,\% for $^{164}$Er, and 0.14\,\% for $^{162}$Er. $^{167}$Er  is the only stable fermonic isotope. The bosonic isotopes have zero nuclear spin ($I=0$) while the fermonic one has $I=7/2$ and shows hyperfine structure. All three electronic states of $^{167}$Er have eight hyperfine levels ranging from $F=J-7/2$ to $F=J+7/2$, where $\vec F=\vec J+\vec I$.  

This paper is structured as follows. Section \ref{AS} describes our experimental methods to investigate the hyperfine structure of the relevant states. Section \ref{TA} reports on our least-squares fitting procedure to obtain the hyperfine constants and isotope shifts from the measured spectra. Section \ref{AI} describes our {\it ab initio} relativistic configuration-interaction  calculations and compares the {\it ab initio} hyperfine constants with the fitted values. We present our conclusions in Sec.\,\ref{concl}. 

\section{Atomic spectroscopy}
\label{AS}

We measure the hyperfine structure of the $^{167}$Er isotope using modulation-transfer spectroscopy \cite{Shirley1982mtp}. The spectroscopy is performed on an atomic Er vapor created with a hollow cathode discharge lamp (HCL). The HCL, based on a sputtering process, has the advantage of providing atomic vapors without the need of a high-temperature atomic source.

We use a commercially available HCL, which is filled with an argon gas at a fixed pressure of $4\,\mathrm{mbar}$ \cite{note1}. By applying a high voltage on the electrodes, the argon gas is ionized and accelerated into the center of the Er-coated cathode. When hitting the surface the kinetic energy of the Ar-ions is high enough to free neutral erbium atoms by sputtering processes \cite{Musha1962csi}. We typically operate the HCL with a voltage of $110\,\mathrm{V}$, giving a discharge current of $9.2\,\mathrm{mA}$. 

\begin{figure}
\includegraphics[width=0.9\columnwidth] {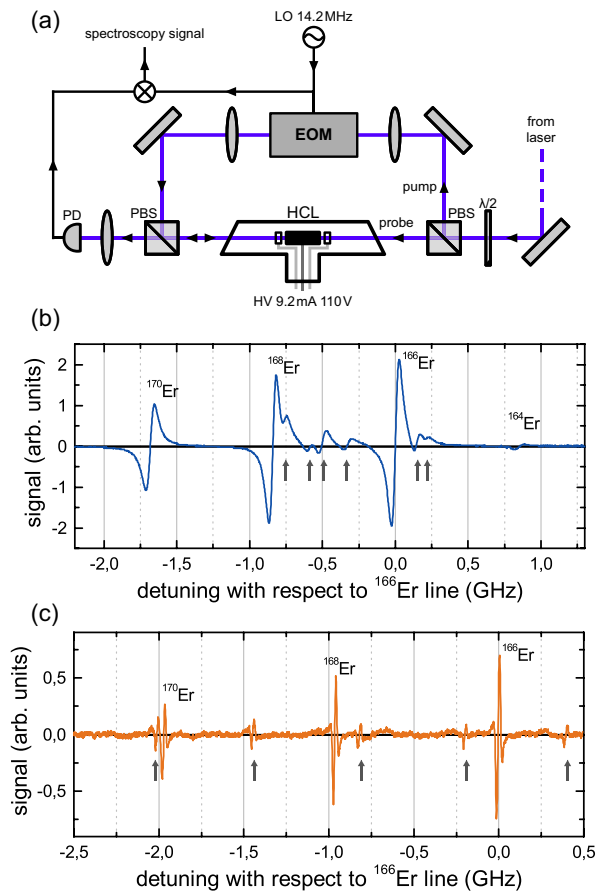}
\caption{(Color online) Modulation transfer spectroscopy of Er for the $401$ nm and $583$ nm transitions. (a) Laser setup for spectroscopy on a hollow cathode discharge lamp (HCL); see text. The pump (probe) light has a power of $3.3\,\mathrm{mW}$ ($0.6\,\mathrm{mW}$) for the $401$ nm transition and $20\,\mathrm{mW}$ ($1\,\mathrm{mW}$) for the $583$ nm transition.
(b), (c) Obtained spectroscopy signals for the $401$ nm and the $583$ nm transitions of different isotopes. Signals related to the hyperfine structure of $^{167}$Er are indicated by arrows. The relative amplitudes of the observed signals reflect the natural isotope abundances.}
\label{fig:expmodspectr}
\end{figure}

We perform a Doppler-free modulation-transfer spectroscopy in the HCL \cite{Lawler1981ahc,Brammer2012dff}. The laser beam is split into a pump and a probe beam, as shown in Fig.\,\ref{fig:expmodspectr}(a). The pump light is modulated with an electro-optical modulator (EOM) driven by a local oscillator (LO) at a frequency of $14.2\,\mathrm{MHz}$ and with a power of $23\,\mathrm{dBm}$ \cite{note2}. A four-wave mixing process transfers the sidebands from the modulated pump beam onto the counter-propagating probe beam \cite{McCarron2008mts}. We acquire the spectroscopy signal by mixing the LO signal with the probe beam signal detected by a photodiode (PD). By setting the LO and the signal either in-phase or shifted by $\pi$, one can obtain a signal proportional to the dispersion or to the absorption of the atomic sample, respectively. In our setup we use the dispersion signal.

In a first set of experiments, we measure the hyperfine structure and the isotope shift of the excited state at $401\,\mathrm{nm}$, with a natural linewidth of $2\pi \times 29.7(5)\,\mathrm{MHz}$ \cite{note3,Mcclelland2006nlo,Lawler2010atp}. A frequency-doubled diode laser is used for the spectroscopy. Figure \ref{fig:expmodspectr}(b) shows the dispersive spectroscopy signal for this transition. The signal is averaged over 16 scans with a scanning speed of $2.4\,\mathrm{GHz/s}$ \cite{note5}.

Our measurement reveals the full hyperfine structure for the fermionic $^{167}$Er. The discussion and the assignment of the observed spectral features are given in Sec.\,\ref{TA}. In addition, we determine the isotope shifts for the bosonic isotopes relative to $^{166}$Er. We measure a shift of $-1681(14)\,\mathrm{MHz}$ for $^{170}$Er, $-840(14)\,\mathrm{MHz}$ for $^{168}$Er, and $+849(17)\,\mathrm{MHz}$ for $^{164}$Er, which is in good agreement with Ref.\,\cite{Connolly2010lsr}. The linewidths are extracted by fitting the derivative of a Lorentzian curve to the data. This gives an averaged value of $2\pi \times 88(8)\,\mathrm{MHz}$, corresponding to about three times the natural linewidth. This broadening of the transition can be explained as a combined effect of collisional and power broadening. For a number density of about $10^{17}\,\mathrm{cm^{-3}}$ and an argon background pressure of $4\,\mathrm{mbar}$, we calculate a collisional broadening of $2\pi \times 8.2\,\mathrm{MHz}$. Considering a total intensity of the pump and probe beams of $ $ $\mathscr{I}=250\,\mathrm{mW/cm^2}$, we estimate a power broadening of a factor of $\sqrt{1+\mathscr{I}/\mathscr{I}_{0}} = 2.3$ with $\mathscr{I}_{0} = 60.3\,\mathrm{mW/cm^2}$ being the saturation intensity. Combining the two contributions, we estimate a broadened linewidth of $2\pi \times 81\,\mathrm{MHz}$, which is in agreement with the observed value. 

In a second set of measurements, we focus on the hyperfine structure of the excited state at $583\,\mathrm{nm}$, with a linewidth of $2\pi \times 186\,\mathrm{kHz}$ \cite{Denhartog2010rlo}. The spectroscopy is performed with a dye laser, which is frequency-stabilized to an ultra-low expansion cavity within 30\,kHz \cite{Frisch2012nlm}. We use a spectroscopy setup similar to the one described above for the $401$ nm transition. Figure \ref{fig:expmodspectr}(c) shows the corresponding spectroscopy signal.

Despite the narrow-line nature of the transition, we could observe five features related to the hyperfine structure of the fermionic isotope and three features for the bosonic ones. The discussion of the hyperfine structure is given in Sec.\,\ref{TA}. We measure an isotope shift of $-975(15)\,\mathrm{MHz}$ for $^{168}$Er and $-1966(14)\,\mathrm{MHz}$ for $^{170}$Er relative to $^{166}$Er, respectively. These values are in good agreement with Ref.\,\cite{Jin1990hsa}.

For this transition, we extract an averaged value for the linewidth of $2\pi \times 23(5)\,\mathrm{MHz}$, corresponding to about 120 times the natural linewidth. This large broadening can again be explained in term of collisional and power broadening. Considering the saturation intensity of $\mathscr{I}_{0}=0.13\,\mathrm{mW/cm^2}$ and our total intensity of $\mathscr{I}=1.3 \times 10^{3}\,\mathrm{mW/cm^2}$, we calculate a power broadening of a factor of 100. Adding the effect of collisional broadening, we obtain an overall linewidth of $2\pi \times 19.3\,\mathrm{MHz}$, which is in agreement with the measured value. Because of this large broadening, we could operate the modulation-transfer spectroscopy at the same LO frequency as the one used for the $401$ nm transition.

\section{Analysis of hyperfine structure}
\label{TA}
\begin{figure}
\includegraphics[scale=0.32]{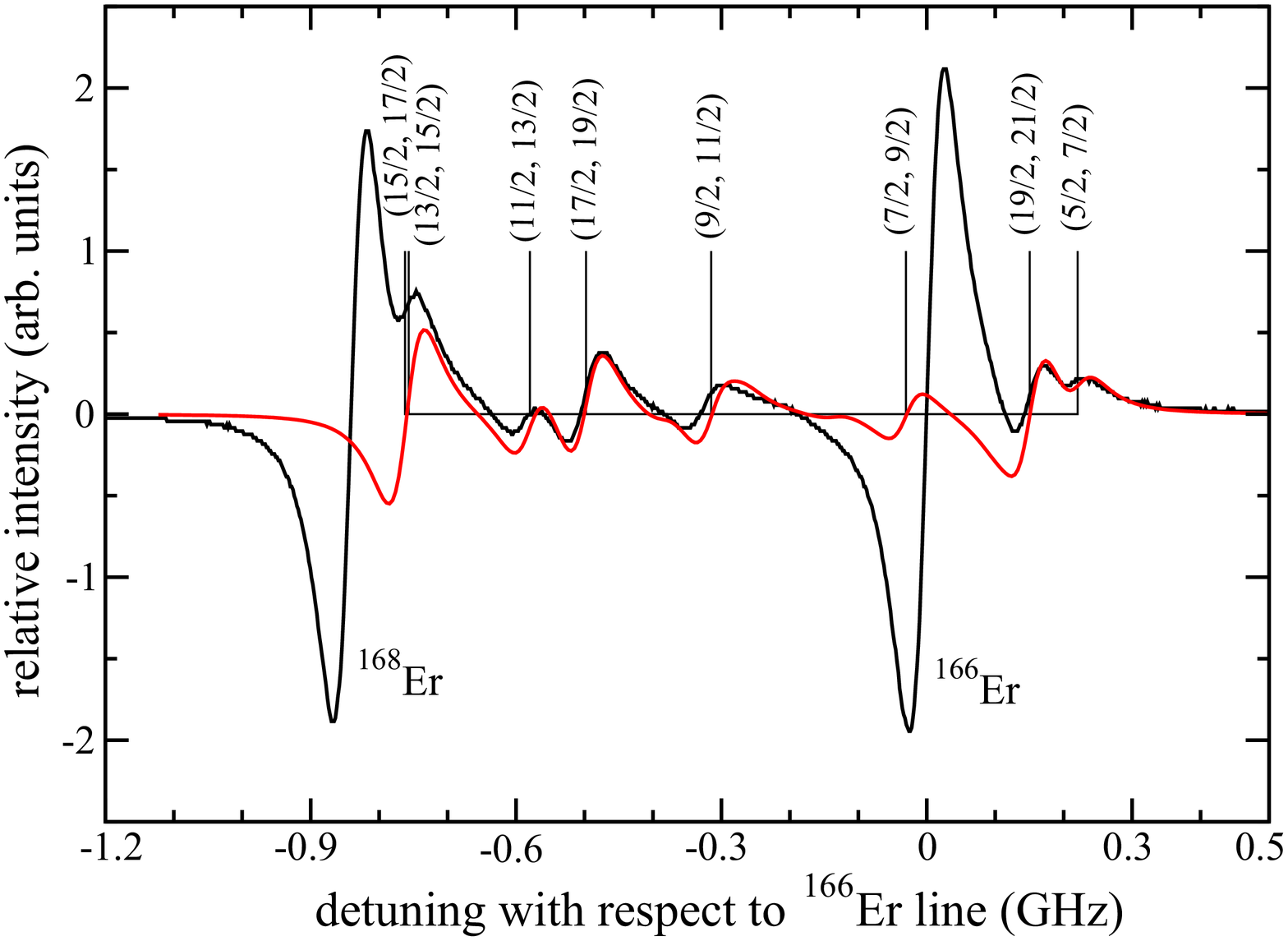}
\caption{(Color online) Spectroscopy signal and hyperfine assignment of the $401$ nm transition of the fermionic $^{167}$Er.
The black solid line is the recorded line. The red line  is a simulated line shape obtained using a nonlinear fit  to the line positions and a linewidth of $\gamma/(2\pi)=90$\,MHz. The simulated line shape is a sum of the first derivative of several Lorentzians, one for each hyperfine transition, whose relative strength is given by a theoretical estimate 
of the line strength. We scaled the overall size of the simulated lineshape to fit to the experiment. The assignment of the P-branch transitions ($F_g \to F_e=F_g+1$) is shown by vertical lines and pairs $(F_g,F_e)$. The hyperfine coefficients of the excited state are $A_e/h =-100.1$\,MHz and $B_e/h = -3079$\,MHz.}
\label{fitted}
\end{figure}

\begin{figure}
\includegraphics[scale=0.32]{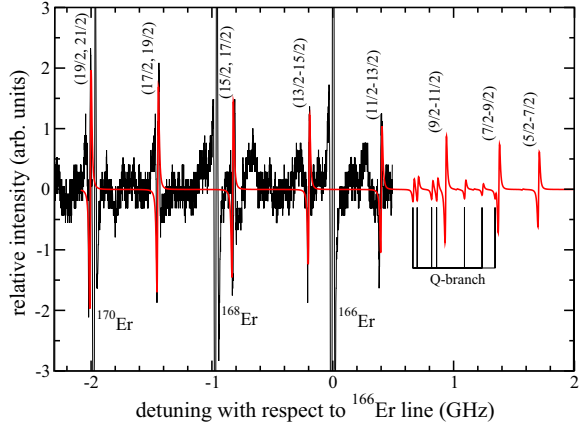}
\caption{(Color online) Spectroscopy signal and hyperfine assignment of the $583$ nm transition of the fermionic $^{167}$Er. The solid black line is the experimental spectrum while the red line is a simulated line shape using a linewidth of $\gamma/(2\pi)=20$\,MHz. The P-branch transitions ($F_g \to F_e=F_g+1$) are assigned by pairs $(F_g,F_e)$. Three P-branch resonances and several Q-branch ($F_g \to F_e=F_g$) resonances are predicted to lie outside of the measurement range. The simulated line-shape is a sum of the first derivative of several Lorentzians, one for each hyperfine transition, whose relative strength is given by a theoretical estimate of the line strength. We scaled the overall size of the simulated line shape to fit to the experiment. The hyperfine coefficients of the excited state are $A_e/h =-172.7$\,MHz and $B_e/h = -4457.2$\,MHz.}
\label{fitted1}
\end{figure}

In this section we describe our fitting procedure to the observed spectra of the five most abundant Er isotopes and we present the resulting hyperfine-structure constants $A_e$ and $B_e$ for $^{167}$Er. The bosonic features are easily assigned as shown in Fig.~\ref{fig:expmodspectr}. The remaining weaker features, which sometimes overlap with those from the bosonic isotopes, are due to $^{167}$Er.

We start with the definition of the transition energies between the ground and an excited state of $^{167}$Er including hyperfine interactions \cite{Arimondo1977}
\begin{equation}
\hbar \omega_{F_eF_g} = \Delta_{167}+\hbar \omega_{166} + E_e(F_e,J_e,I) - E_g(F_g,J_g,I)\,,
\end{equation} 
where $\Delta_{167}$ is the $^{167}$Er  isotope shift relative to the
transition energy $\hbar \omega_{166}$ of the bosonic $^{166}$Er atom, the most abundant isotope, and  $E_e(F_e,J_e,I)$
and $E_g(F_g,J_g,I)$ are hyperfine energies of the excited and ground
state, respectively.  The quantum numbers $F_i$ and $J_i$ with $i=e$ or $g$ are the
total atomic and electronic angular momentum of the excited and ground
state, respectively, and
\begin{eqnarray}
\lefteqn{E_i(F_i,J_i,I) = \frac{1}{2}A_i\,C_i  }\label{hfs}\\
   &&  \quad\quad +\frac{1}{2}B_i\frac{3C_i(C_i+1)-4I(I+1)J_i(J_i+1)}{2I(2I-1)2J_i(2J_i-1)}\,,
    \nonumber
\end{eqnarray}
where $C_i=F_i(F_i+1)-J_i(J_i+1)-I(I+1)$.
Finally, the transition energies $\Delta_{A}+\hbar \omega_{166}$ define the isotope shift for bosonic Er isotopes with atomic number $A$.

In addition to the resonance positions, we can calculate the line shape of the fermionic spectral features $S(\omega)$ by noting that the signal is well approximated by
\begin{equation}
  S(\omega) \propto -\sum_{F_e F_g} Q_{F_e,F_g}  \frac{d}{d\omega} {\cal L}(\omega-\omega_{F_eF_g},\gamma)
\end{equation}
as a function of laser frequency $\omega$, where the sum is over all $(F_g,F_e)$ hyperfine lines, and ${\cal L}(\omega,\gamma)$ is a Lorentzian centered around zero with linewidth $\gamma$  \cite{note4}. Consequently, for an isolated line the resonance occurs when the signal is zero. The fluorescence line strength $Q_{F_e,F_g}$ is
\begin{eqnarray}   
Q_{F_e,F_g} &=& 
          \sum_{M_eM_g q}    |\langle (J_gI)F_gM_g | d_{1q} | (J_eI)F_eM_e\rangle|^2
       \nonumber   \\ 
       & =&
         {\hat F}_g {\hat F}_e {\hat J}_g  \left(\begin{array}{ccc}F_g &F_e &1\\J_e &J_g& I \end{array}\right)^2 
                               |\langle J_g || d || J_e\rangle|^2\,, 
 \end{eqnarray}
where the $M_i$ are magnetic quantum numbers, $d_{1q}$ is the electric dipole-moment operator, and  we have assumed equal population for all hyperfine states $F_eM_e$ of the electronic excited state. Finally, $\hat F=2F+1$, $({\cdots\atop\cdots})$ is a six-$j$ symbol, and $\langle J_g || d || J_e\rangle$ is a reduced dipole matrix element independent of  $F_g$ and $F_e$.

We use a nonlinear least-squares fit to the experimental spectra  to determine the hyperfine constants and isotope shift $\Delta_{167}$ of the excited states.  The fit is based on six resolved hyperfine features for the $401$ nm line and five resolved features for the narrow $583$ nm line. In our analysis, we hold the hyperfine constants for the ground state to the literature values of $A_g/h= -120.487(1)$ MHz and $B_g/h=-4552.984(10)$ MHz \cite{Childs1983}, which have significantly lower uncertainties than those for the excited states.

\begin{figure}
\includegraphics[scale=0.30]{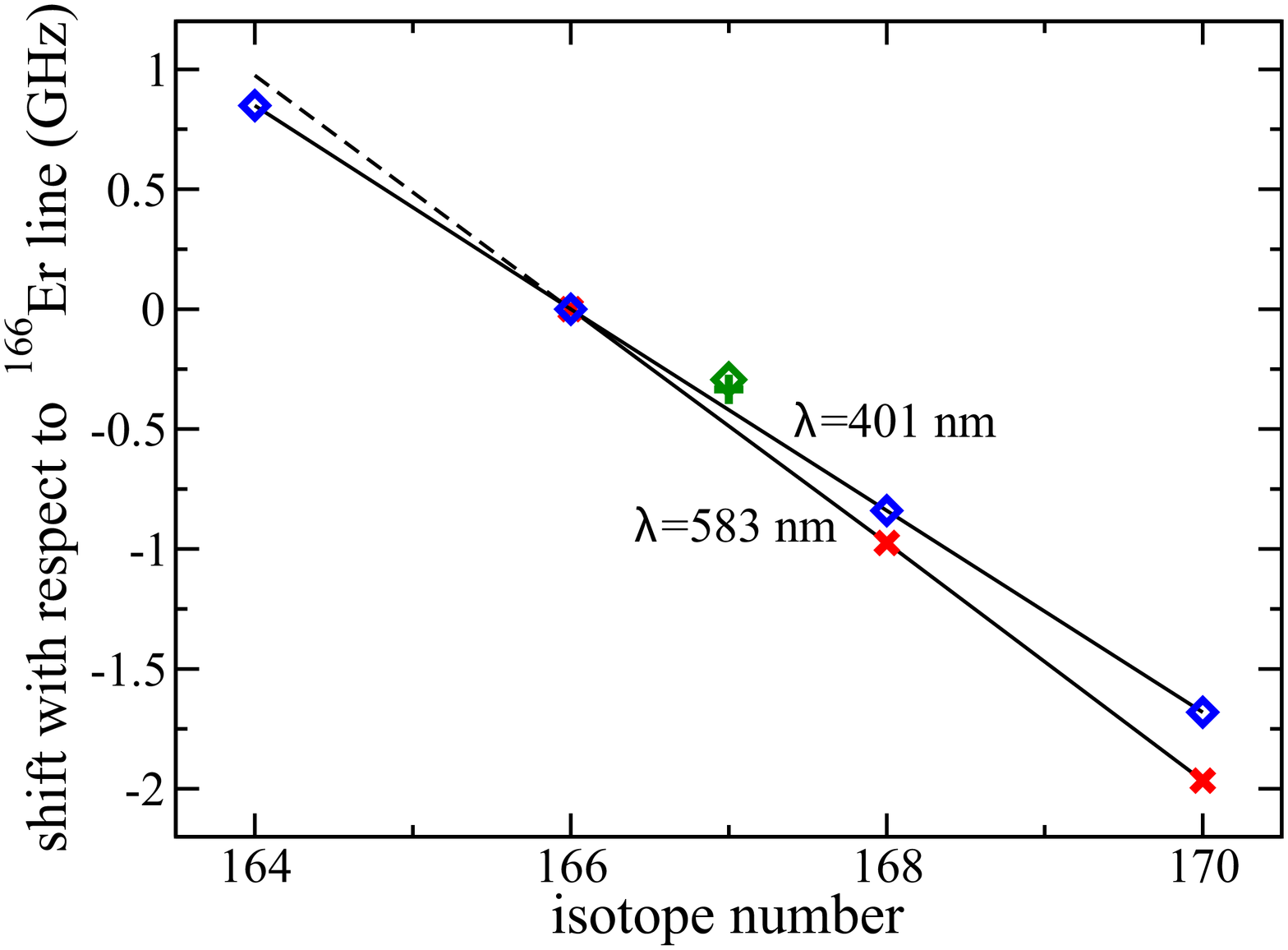}
\caption{(color online). Isotope shifts for the $583$ nm and $401$ nm lines of the isotopes $^{164}$Er up to $^{170}$Er as a function of mass number where the transition energy for the bosonic  isotope $^{166}$Er is taken as energy reference. The isotope shift of the bosonic isotopes falls on a single straight line, with the isotope shift of the center of gravity of the fermionic $^{167}$Er isotope, green cross and green square, is slightly displaced from this linear dependence.}
\label{isotope}
\end{figure}

Figure~\ref{fitted} and \ref{fitted1} are the results of our fit for the $401$ nm and $583$ nm line, respectively. We observe remarkable agreements between the simulated and experimental spectra.  For the excited $401$ nm level, we extract the best value for the hyperfine coefficients to be $A_e/h =-100.1(3)$\,MHz and $B_e/h = -3079(30)$\,MHz. Using these coefficients and those for the ground state,  we obtain resonance positions that  agree to better than 11 MHz with the experimental values.  For the excited $583$ nm level, we fit the line shape of the spectral features while the resonance positions are calculated by using the hyperfine constants of the excited states,  $A_e/h =-172.7$\,MHz and $B_e/h = -4457.2$\,MHz,  from Ref.\,\cite{Jin1990hsa}. We note that the additional structure in the experimental data, which is not fitting to the theoretical curve, originates from a slightly misadjusted phase of the local oscillator in the spectroscopy setup. Table~\ref{energies} compares the theoretical and experimental hyperfine energies  $\Delta_{167}+E_e(F_e,J_e,I)+E_e(F_g,J_g,I)$ for the $583$ nm and $401$ nm transitions in $^{167}$Er and lists the corresponding quantum numbers of $F_g$ and $F_e$.   Table \ref{isotopeshifts} and Fig.\,\ref{isotope} show the resulting isotope shifts $\Delta_{A}$ as a function of the mass number $A$ relative to the energy of the  $^{166}$Er isotope.

\begin{table}[b]
\caption{The observed and calculated hyperfine energies $\Delta_{167}+E_e(F_e,J_e,I)+E_e(F_g,J_g,I)$ for the $583$ nm and $401$ nm lines in $^{167}$Er. The theoretical values are based on the hyperfine coefficients $A_e/h = -172.7$\,MHz and $B_e/h = -4457.2$\,MHz \cite{Jin1990hsa} for the $583$ nm line and our values $A_e/h =-100.1$\,MHz and $B_e/h = -3079$\,MHz for the $401$ nm line.} 
\begin{ruledtabular}
\begin{tabular}{ccl|ccl}
       Obs.     &  Calc.     & $(F_g,F_e)$ & Obs.      &  Calc.    & $(F_g,F_e)$     \\ 
        Energy     &   Energy    &  & Energy     &   Energy    &       \\ 
       (MHz)                    &  (MHz)              &              & (MHz)                    &  (MHz)              &               \\
       \hline
\multicolumn{3}{c|}{ 583~nm }&  \multicolumn{3}{c}{401~nm } \\ \hline 
   $-$2011               &  $-$2011       &  (19/2, 21/2)&  $-$761           & $-$762         & (15/2, 17/2)  \\
   $-$1449               &  $-$1454       &  (17/2, 19/2)&   -            & $-$757         & (13/2, 15/2)  \\
   $-$820                &   $-$834       &  (15/2, 17/2)& $-$589           & $-$580         & (11/2, 13/2)    \\
   $-$200                &   $-$203       &  (13/2, 15/2)&  $-$498           & $-$498         & (17/2, 19/2)   \\
    393                &    396       &  (11/2, 13/2)&  $-$325           & $-$315         & (9/2, 11/2) \\
     -                 &     941       &  (9/2, 11/2) &    -            & $-$31          & (7/2, 9/2)   \\
     -                 &     1369      &  (7/2, 9/2)  &   150           &  150         & (19/2, 21/2)   \\
     -                 &     1709      &  (5/2, 7/2)  &  220           &  220         & (5/2, 7/2)   \\
\end{tabular}
\end{ruledtabular}
\label{energies}
\end{table}

\begin{table}[b]
\caption{Observed isotope shift of the Er isotopes for the $583$ nm and $401$ nm lines. The transition energy for the bosonic isotope $^{166}$Er is taken as energy reference. The isotope shift of the center of gravity of fermionic $^{167}$Er was obtained from fitting.
} 
\begin{ruledtabular}
\begin{tabular}{c|cc|cc}
       Isotope &Obs.     & Calc. & Obs.    &  Calc.  \\ 
        & Energy    &  Energy &  Energy     &   Energy \\ 
                              &            (MHz)                     &  (MHz)                  &               (MHz)                     &  (MHz)                          \\
       \hline
& \multicolumn{2}{c|}{583~nm }&  \multicolumn{2}{c}{401~nm } \\ \hline 
   170                &  $-$1966(14)       &                &  $-$1681(14)          &                                   \\
    168               &   $-$975(15)        &                & $-$840(14)           &                                     \\
    167               &                    &  $-$337(11)      &                    &   $-$297(6)                   \\
    166                &    0             &                &   0               &                                    \\
    164                &     -      &                 &  $+$849(17)           &                                 \\
\end{tabular}
\end{ruledtabular}
\label{isotopeshifts}
\end{table}

\section{{\it ab initio} hyperfine constants} \label{AI}

In conjunction with the experimental measurements and fits, we have performed extensive {\it ab~initio} electronic structure calculations of the magnetic dipole $A$ and electric quadrupole $B$ hyperfine constants. They describe the coupling of the nuclear spin ${\vec I}$ to the total electron angular momentum ${\vec J}$, due to the magnetic dipole and electric quadrupole interaction, respectively. The latter originates from the electric field gradient created by the electrons at the nuclear location. We were interested to reproduce the known constants for the Er ground state as well as those of the excited state at  the $583$ nm line obtained by Ref.~\cite{Childs1983,Jin1990hsa}. We can then confirm our  measurement of the unknown constants of the excited level at the $401$ nm line.

The {\it ab~initio} calculations of the hyperfine structure constants have been performed using a relativistic multiconfiguration Dirac-Fock (MCDF) method \cite{Kotochigova1998}. In this method we perform an all-electron calculation of the wave function leading to an accurate description of the electron-spin density near the nucleus. The eigenfunctions  are superpositions of non orthogonal many-electron determinants of one-electron Dirac-Fock functions for the core and valence orbitals  and Sturm functions for virtual orbitals.  Both types of one-electron orbitals are optimized for either the $4f^{12}6s^2$ ground or $4f^{12}6s6p$ excited-state reference configurations.

The hyperfine splittings of atomic levels are due to interactions between electrons and nuclear multipole moments.  In the configuration interaction picture and using atomic units the  $A$ and $B$  constants are given by 
\begin{eqnarray} 
A(J)& =& \frac{g_I
\mu_N }{M_J} \langle \Psi, JM_J|\sum_i \frac{[\vec{r}_i \times
\vec{\alpha}_i]_{00}}{r_i^3} | \Psi, JM_J\rangle\,,\\ B(J) & =&
\frac{2Q}{M_J} \sqrt{\frac{2J(2J-1)(2J+1)}{(2J+2)(2J+3)}}  \\ &&\quad\quad
\times\langle \Psi, JM_J|\sum_i \frac{Y_{20}(\hat{r}_i)}{r^3_i}|\Psi,
JM_J\rangle  \,, \nonumber 
\end{eqnarray} 
where the sum $i$ is over all electrons with positions $\vec{r}_i$ with respect to the nucleus,
$Y_{\ell m}(\hat{r})$ are spherical harmonics, and ${\vec\alpha}_i$ is the Dirac matrix for electron $i$. Furthermore, $g_I$ is the nuclear g-factor, $\mu_N$ is the nuclear magneton in atomic units, and  $Q$ is the nuclear quadrupole moment.  The relativistic electronic eigenfunctions $|\Psi, JM_J \rangle=\sum_\beta c_\beta |\phi_\beta, JM_J\rangle$, obtained from
the  configuration-interaction calculations with relativistic determinants $|\phi_\beta, JM_J\rangle$ and CI coefficients $c_{\beta}$, have  total angular momentum  $J$ and projection $M_J$.

When an atom has open or unfilled electron shells, it leads to an unbalanced electron-spin density near the location of the nucleus. As hyperfine constants are proportional to the difference in electron-spin densities this leads to nonzero  $A$ and $B$ coefficients. To
account for this effect we use a model, where the single-electron orbitals differ for each spin direction or more precisely for each spinor of the Dirac-Fock equation. Alternatively,
this implies different exchange potentials for electrons with spin up or down. 

We use three restricted active spaces (RAS) to classify the electron Dirac-Fock and Sturm orbitals, ensuring an efficient and compact CI expansion that, nevertheless, remains accurate. The first group of orbitals, RAS1, contains the occupied spinors of the relevant reference configuration. We have studied convergence of the hyperfine structure constants as the active set of orbitals was systematically increased. For our most-precise Er atom calculation RAS1
includes the occupied $1s^2, 2s^2, 2p_{1/2}^2, 2p_{3/2}^4, ...$, and $4f_{5/2}^6$ shell electrons. We allow up to one electron to be excited out of RAS1 into the two other active spaces. The second group, RAS2, contains the open-shell $4f_{7/2}^6$, $6s_{1/2}^2$, and $6p_{1/2}, 6p_{3/2}$ spinors, while the third group, RAS3, contains spinors that are unoccupied in the reference configuration.  These latter virtual Sturm orbitals are the high-lying $s$-wave spinors from $7s$ up to $13s$, $p$-wave spinors from $7p$ up to $11p$, and the $5d$-spinor. For both RAS2 and RAS3 we allow up to two electrons to enter or leave.

With this basis our finite-nuclear-size and finite-nuclear mass corrected {\it ab~initio} values of the $A$ and $B$ hyperfine constants are $-120.42$ MHz  and $-4554$ MHz for the ground state level, $-174$ MHz and $-4057$ MHz for the excited  level at  583 nm, and $-100$ MHz and $-3424$ MHz for the excited level at 401 nm, respectively.  Consequently, the {\it ab~initio} $A$ constants agree with experimentally determined values to better than 1\%, whereas the $B$ constants differ by up to 11\% for the two excited states.  For the ground state the agreement for the $B$ constant is also better than 1\%.

\section{Conclusions}
\label{concl}
\begin{table}[t]
\caption{A summary of the relevant hyperfine A and B constants for ground  and excited states (e.s.) usable for laser-cooling of $^{167}$Er.}
\begin{ruledtabular} 
\begin{tabular}{ccccc}
       state & $J$ & $A/h$ (MHz) & $B/h$ (MHz) & Ref.  \\ 
       \hline
   ground state   & 6 &  $-$120.487(1) &  $-$4552.984(10) &  \cite{Childs1983}  \\  
    {\it ab~initio}  &  &  $-$120.42 &  $-$4554            &  this work  \\ 
       \hline
   583-nm e.s.    & 7 & $-$172.70(7)   &  $-$4457.2(29)       &   \cite{Jin1990hsa} \\      
    {\it ab~initio}  &  & $-$174      &  $-$4057        &   this work \\  
     \hline
   401-nm e.s.    & 7 & $-$100.1(3)  &  $-$3079(30)         &  this work  \\
    {\it ab~initio}  &  & $-$100       &  $-$3424       &   this work \\ 
\end{tabular}
\end{ruledtabular}
\label{tab:hfconstants}
\end{table}

\begin{figure}[b]
\includegraphics[width=0.85\columnwidth] {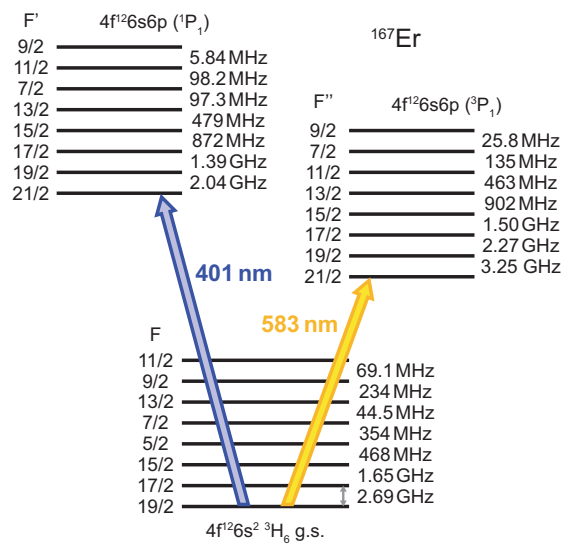}
\caption{(Color online) Hyperfine levels of the ground (g.s.) and two excited states of $^{167}$Er with particular interest for laser cooling. The level splitting was calculated using A and B constants given in Table \ref{tab:hfconstants} for the respective transitions. The arrows depict two laser-cooling transitions. The transition at 401\,nm used for Zeeman slowing is shown in blue and the transition used for magneto-optical trapping at 583\,nm is shown in yellow.}
\label{fig:conclhfstruct}
\end{figure}

We have used laser modulation-transfer spectroscopy on atomic Er as well as performed {\it ab initio} electronic structure calculations of Er to obtain the magnetic dipole and electric quadrupole constants for the only stable fermionic isotope, $^{167}$Er. We focused on transitions from the $4f^{12}6s^2$ ($J=6$) ground state to two $J=7$ levels within the excited $4f^{12}6s6p$ configuration. A least-squares algorithm applied to the experimentally-measured hyperfine-structure energies gives accurate values for the two constants as well as values for the isotope shift of five isotopes.  The {\it ab initio} calculation is based on a multiconfiguration Dirac-Fock method where we allow no more than two electrons to be excited from and between the active spaces. The method has no further adjustable parameters.

Our results  are summarized in Table~\ref{tab:hfconstants} and Fig.~\ref{fig:conclhfstruct}. We find that the {\it ab initio} $A$ coefficients for all three states and the $B$ coefficient for the ground state agree to better than 1\% with the experimental values, which is in a surprisingly good agreement considering the complex electron-shell structure of the Er atom. We note that the  {\it ab~initio} electric quadrupole constants $B$ for the two excited states exhibit a larger deviation from the experimental values. This might be a consequence of  missing key configurations: the excited states have three open shells, $4f^{12}$, $6s$, and $6p$, from which more than two electrons might need to be excited. In addition, Sternheimer shielding (e.g., distortions in the electron shells by the nuclear quadrupole moment), which is not considered in our MCDF theory, might cause significant corrections.

\subsection*{Acknowledgments}
We would like to acknowledge R. Grimm for fruitful discussions and A. Rietzler, S. Baier, and M. Springer for technical support. The Innsbruck team acknowledge support by the Austrian Ministry of Science and Research (BMWF) and the Austrian Science Fund (FWF) through a START grant under Project No.\,Y479-N20 and by the European Research Council under Project No.\,259435. C.B. and S.K. acknowledge support by the AFOSR Grant No.~FA 9550-11-1-0243 and ARO MURI Grant No.~W911NF-12-1-0476. They also thank Dr. I. I. Tupitsyn for useful discussions.


\begin{thebibliography}{46}
\bibitem[{not({\natexlab{a}})}]{note2}
\bibinfo{note}{The modulation frequency is chosen such that the steepness of the spectroscopy signal is maximized, according to \cite{McCarron2008mts}.}
\bibitem[{not({\natexlab{a}})}]{note5}
\bibinfo{note}{The frequency is monitored during recording the spectroscopy signal and the systematic uncertainty due to nonlinear scanning is estimated to be $12\,\mathrm{MHz}$.}
\bibitem[{not({\natexlab{a}})}]{note4}
\bibinfo{note}{Pressure broadening, giving a Gaussian contribution to the full profile, is small and can thus be neglected.}
\end{thebibliography}
\end{document}